\documentclass[]{spie}  
\usepackage[latin1]{inputenc}
\usepackage[]{graphicx}

\newcommand{\degr}{\ensuremath{^{\circ}}}

\title{First astrophysical results from AMBER/VLTI}

\author{F. Malbet\supit{a}, R.G. Petrov\supit{b}, G. Weigelt\supit{c},
  P. Stee\supit{d}, E. Tatulli\supit{e}, A. Domiciano de
  Souza\supit{b,d}, F. Millour\supit{a,b} and the AMBER consortium
\skiplinehalf
\supit{a}Laboratoire d'Astrophysique de Grenoble, BP 53, F-38041
  Grenoble cedex 9, France; \\
\supit{b}Laboratoire Universitaire d'Astrophysique de Nice, France;\\
\supit{c}Max-Planck Institut für Astrophysick, Bonn, Germany;\\
\supit{d}Observatoire de la Côte d'Azur, Nice, France; \\
\supit{e}Osservatorio Astrofisico di Arcetri, Firenze, Italy}

\authorinfo{Further author information: (Send correspondence to
  F.~Malbet)\\
  E-mail: fabien.malbet@obs.ujf-grenoble.fr}

\begin{document} 
  \maketitle 

\begin{abstract}
  The AMBER instrument installed at the Very Large Telescope (VLT)
  combines three beams from as many telescopes to produce spectrally
  dispersed fringes with milli-arcsecond angular scales in the near
  infrared. Two years after installation, first scientific
  observations have been carried out during the \emph{Science
    Demonstration Time} and the \emph{Guaranteed Time} mostly on
  bright sources due to some VLTI limitations.  In this paper, we
  review these first astrophysical results and we show which types of
  completely new information is made available by AMBER.

  The first astrophysical results have been mainly focusing on
  stellar wind structure, kinematics, and its interaction with dust
  usually concentrated in a disk. Because AMBER has dramatically
  increased the number of measures per baseline, this instrument
  brings strong constraints on morphology and models despite a
  relatively poor $(u, v)$ coverage for each object.
\end{abstract}



\section{INTRODUCTION}
\label{sect:intro}  

AMBER is one of the first-generation instruments of the \emph{Very
  Large Telescope Interferometer (VLTI)} that has been described by
Petrov et al.\cite{2003SPIE.4838..924P,2003Ap&SS.286...57P}. The
science program prepared has already partly been
described by Malbet et
al.\cite{2003SPIE.4838..917M,2004SPIE.5491.1722M} in previous SPIE
papers.

AMBER is an interferometric beam combiner for the VLTI working in the
near-infrared $J$, $H$, $K$ bands. It is able to simultaneously handle
3 beams coming from 3 identical telescopes. AMBER interferograms are
spectrally dispersed with a resolution of about 35, 1200 and 10000.
Therefore AMBER can measure vibilities and a closure phase in a few
hundred different spectral channels.

The spectral coverage, the spectral resolution and the better
sensitivity compared to small-aperture interferometers give access to
many new astrophysical fields that we describe preliminaryly in this
review.  

\section{Wind/disk connection in young stars of intermediate
  masses}

\begin{figure}[t]
  \centering
  \includegraphics[width=0.6\hsize]{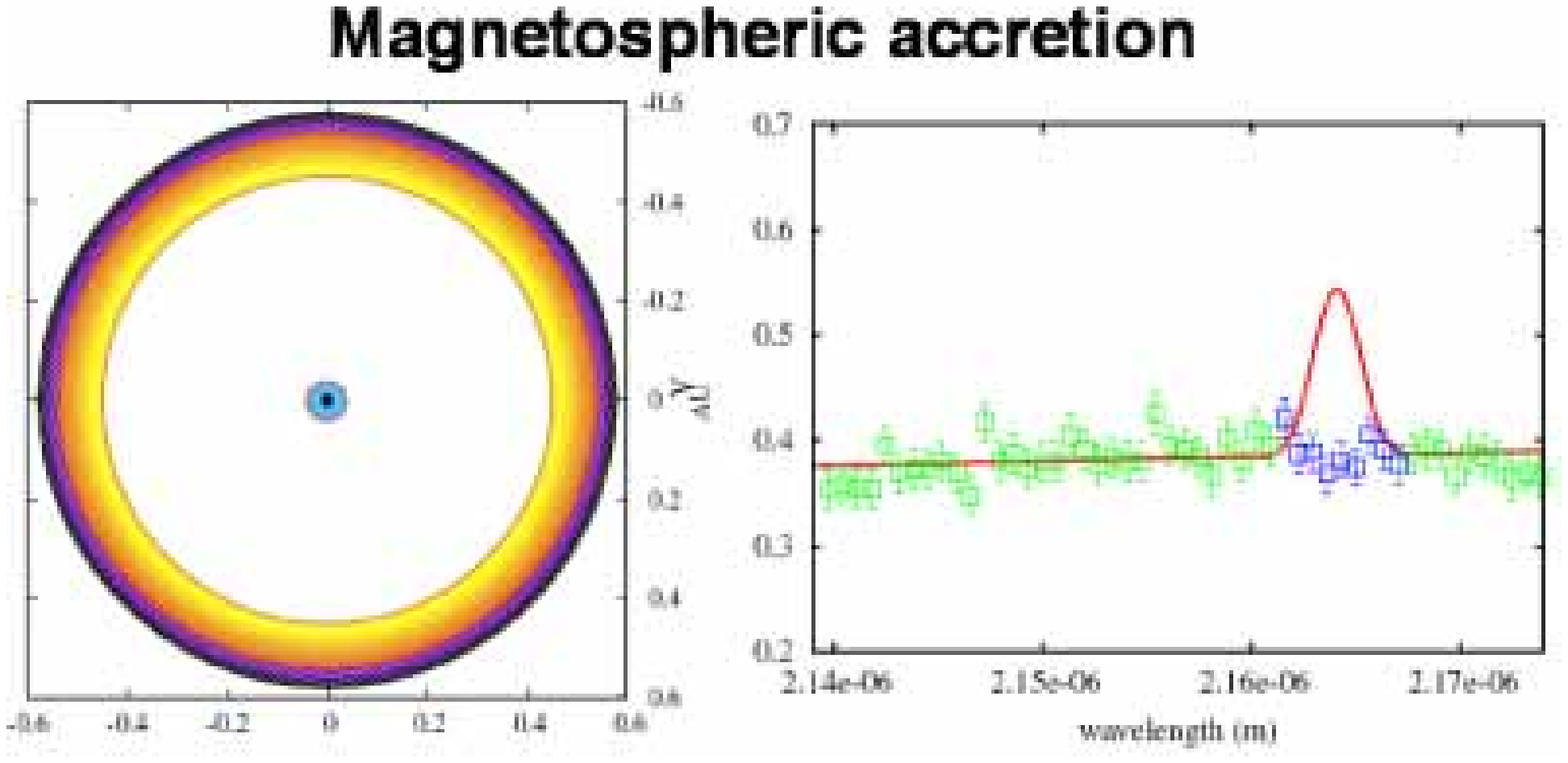}
  \includegraphics[width=0.6\hsize]{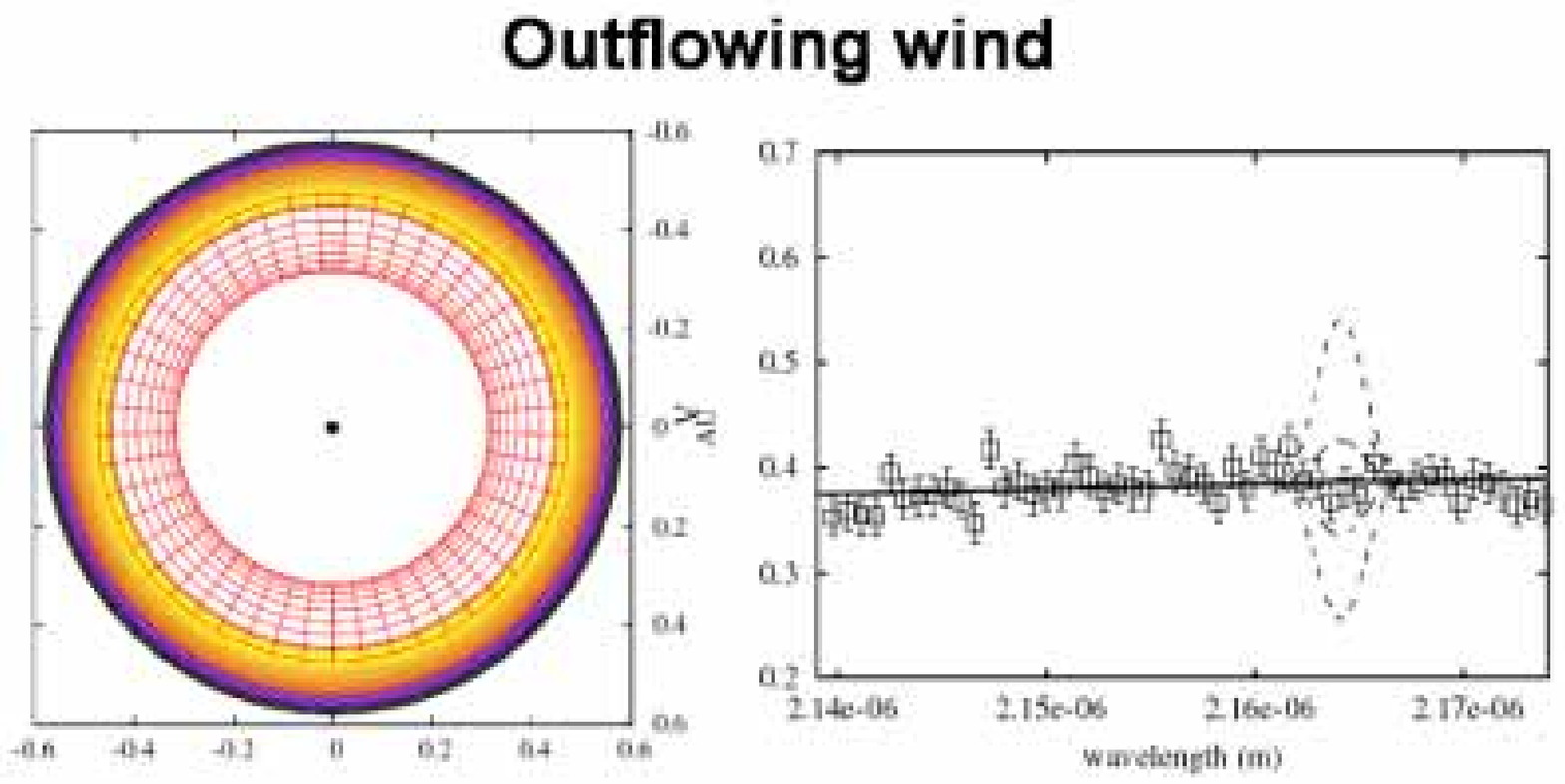}
  \caption{Model (left part) and observation (right part) of the young
    stellar object HD~104237. The upper panel shows, in solid line, the
    result of a simulation of the gas emission if it would come from the
    magnetosphere. The bottom panel shows in solid line the result
    from a simulation of an outflowing wind. The dashed line corresponds
    to different sizes of the wind launching region. The best fit
    corresponds to a launching region between 0.3 and 0.5AU.}
  \label{fig:hd104237}
\end{figure}

The young stellar object MWC~297 is an embedded Herbig Be star
exhibiting strong hydrogen emission lines and a strong near-infrared
continuum excess which has been observed with the AMBER instrument during
its first commissioning run\cite{2005astro.ph.10350M}.  MWC~297 has
been spatially resolved in the continuum with a visibility of $\sim
0.50$ as well as in the Br$\gamma$ emission line where the visibility
decreases to a lower value of $\sim 0.33$. A first interpretation of
this result is that the gas emitting the Br$\gamma$ emission line
is located in a region slightly larger than the one from which the
dust continuum emission arises.

A picture emerges in which MWC~297 is surrounded by an equatorial flat
disk that is possibly still accreting and by an outflowing wind.  
AMBER's unique capability to measure spectral visibilities allowed
Malbet et al.\cite{2005astro.ph.10350M} for the first time to compare
the apparent geometry of a wind with the disk structure in a young
stellar system. However the measurements were not sufficient to
unambiguously choose between the different scenarios of wind-launching
region: disk wind or X-wind.

A lower mass, less active system, the Herbig Ae system HD104237 has
also been observed with AMBER\cite{2006tatulli}. The central A2 emission
line star is surrounded by a circumstellar disk, which causes the
infrared excess emission and drives a jet seen in Ly-$\alpha$ images.
The optical spectrum shows a rather narrow H$\alpha$ emission with a
P-Cygni profile.  The visibility measured by AMBER does not vary
between the continuum and the Br$\alpha$ line, even though the line is
strongly detected in the spectrum, with a peak intensity 35\% above
the continuum (see Fig.~\ref{fig:hd104237}). This demonstrates that
the line and continuum emission have similar size scales. Assuming
that the K-band continuum excess originates in a puffed-up inner rim
of the circumstellar disk, Tatulli et al.\cite{2006tatulli} conclude that
this emission most likely arises from a compact disk wind very close
to the inner rim location.

These two results show that AMBER on the VLTI is going to be a
major tool for understanding the very close environment of young stars
and will disentangle the region of emission from dust and gas,
especillay coming from the disk and the wind.

\section{Rotating gas envelope around hot Be and B[e] stars}

\begin{figure}[t]
  \begin{center}
    \includegraphics[width=0.5\hsize]{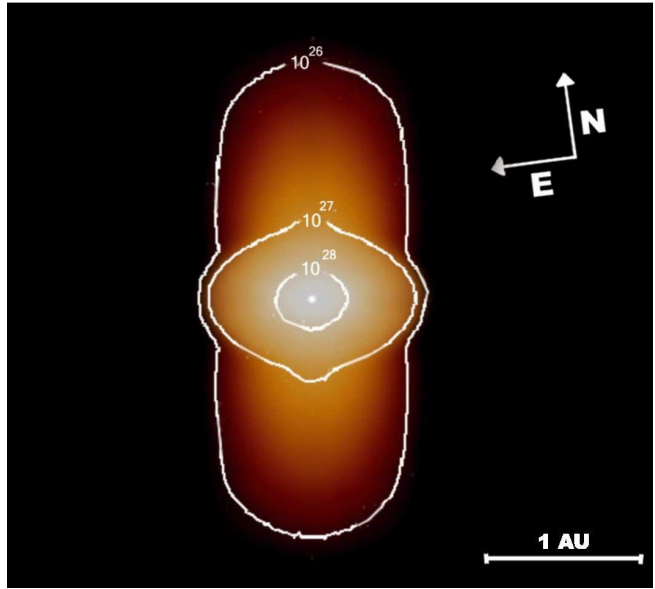}
    \caption{Intensity map in the continuum at 2.15 $\mu$m obtained
      with SIMECA for the best model parameters of $\alpha$ Arae.
      The inclination angle is 55\degr, the central bright region is
      the flux contribution from the thin equatorial disk whereas the
      smoother regions originate from the stellar wind. The brightness
      contrast between the disk and the wind is globally $\sim$ 30 but
      can reach 100 if you compare the inner region of the disk with
      the outer parts of the wind.}
    \label{fig:map_continu}
    \end{center}
\end{figure}

Several emission-line supergiants have been scrutinized by AMBER:
$\alpha$ Arae, one of the closest Be stars\cite{2006astro.ph..6404M},
$\kappa$ Canis Majoris, one of the brightest ones\cite{2006meilland} and
CPD-57{\degr}2874 a B[e] supergiant star characterized by dust
emission\cite{2005astro.ph.10735D}.

\begin{figure}[t]
  \begin{center}
    \includegraphics[width=0.45\hsize]{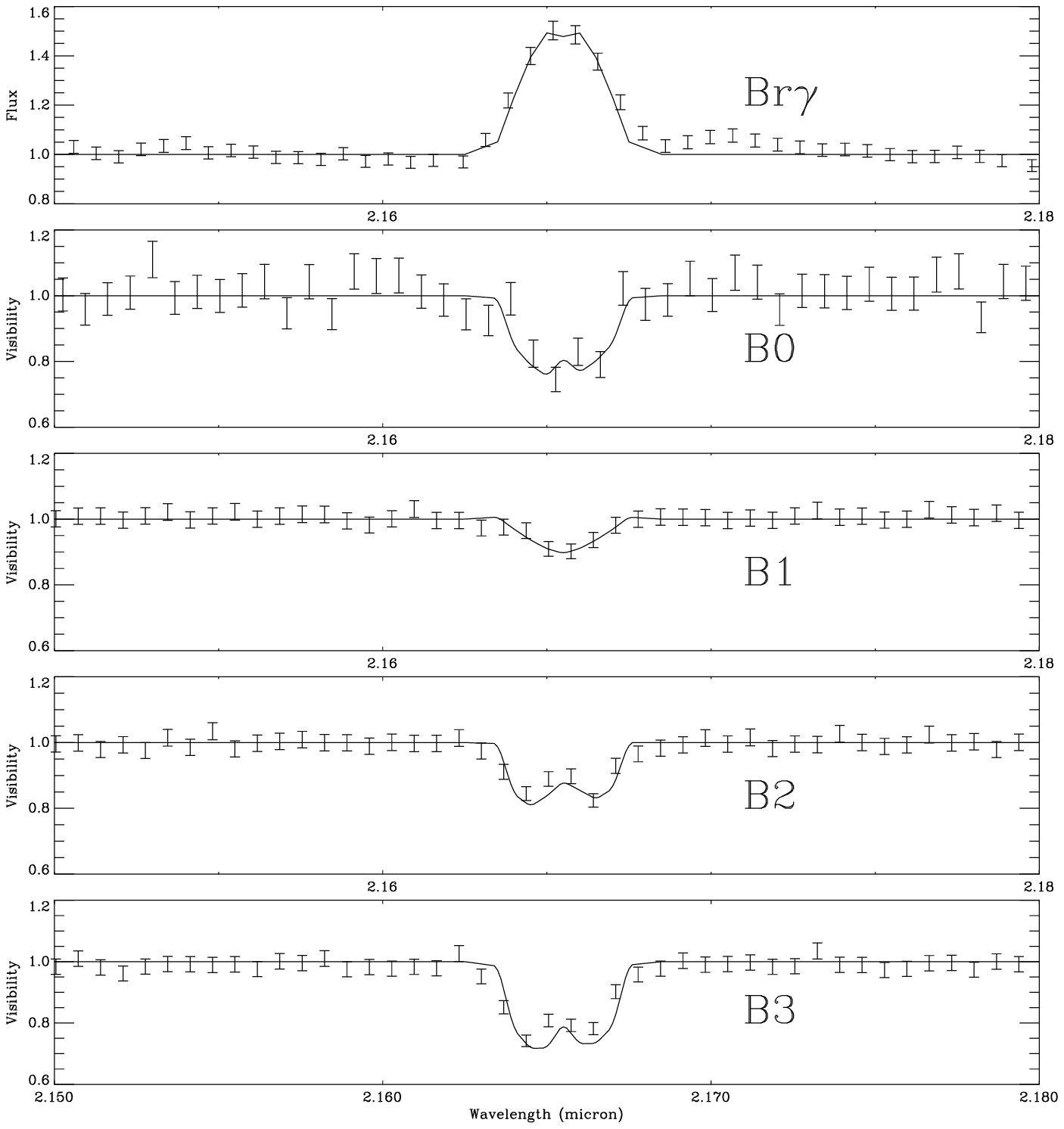}\hfill
    \includegraphics[width=0.47\hsize]{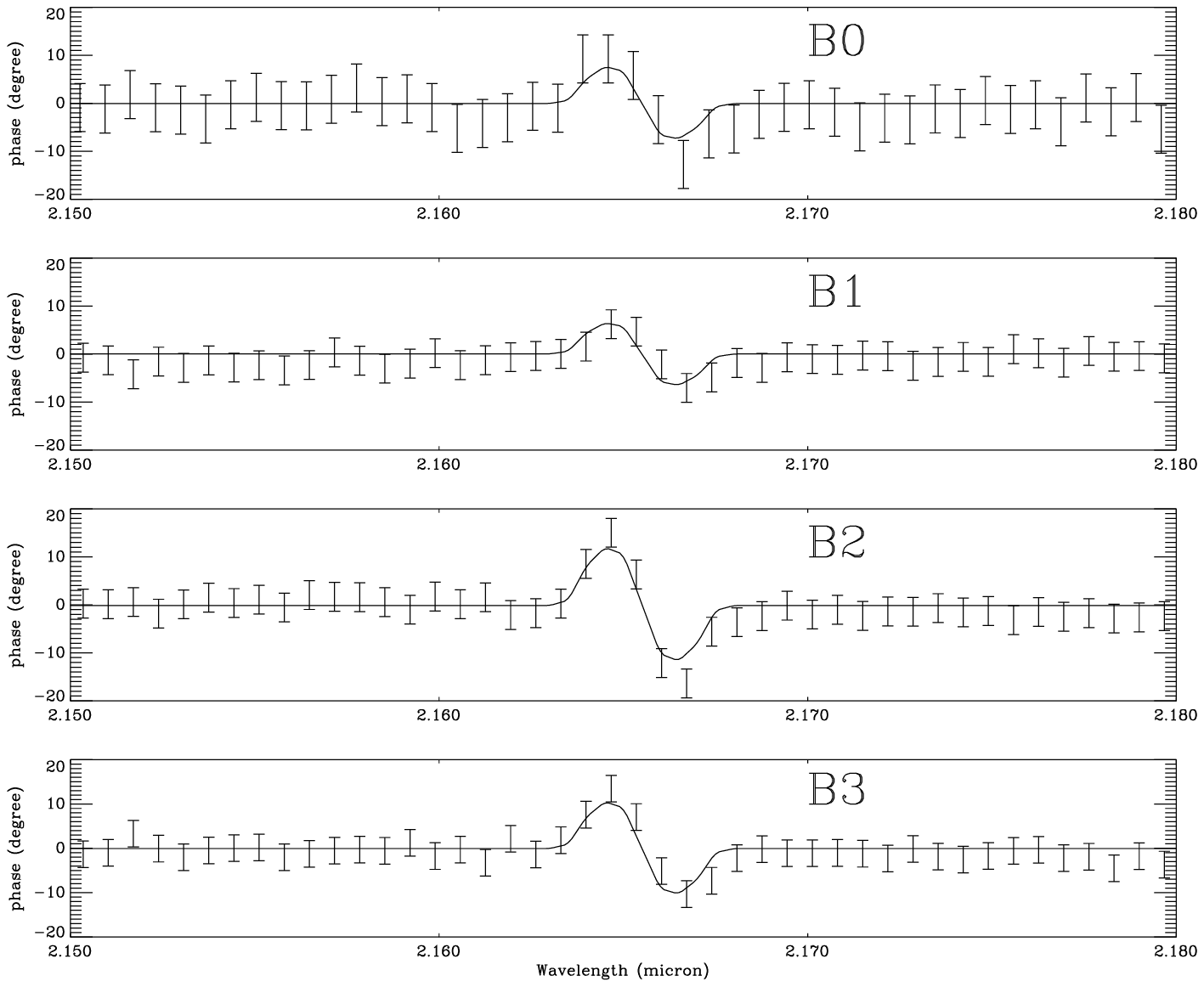}
    \caption{Relative visibility (left) and differential phases
      (right) of $\alpha$ Arae across the Br$\gamma$ line profile for
      several VLTI baselines. The first picture from the top left is
      the Br$\gamma$ line profile. The plain line are the fits we
      obtain with SIMECA from our best model whereas the VLTI/AMBER
      data are the points with error bars.}
    \label{fig:alfara}
    \end{center}
\end{figure}

The AMBER instrument operating in the $K$ band provides a gain by a
factor of 5 in spatial resolution compared to previous VLTI/MIDI
observations of $\alpha$ Arae. Moreover, it is possible to combine the
high angular resolution provided with the (medium) spectral resolution
of AMBER to study the kinematics of the inner part of the disk and to
infer its rotation law (see Fig.~\ref{fig:alfara}). Meilland et
al.\cite{2006astro.ph..6404M} obtained, for the first time, the direct
evidence that the disk is in Keplerian rotation, answering a question
that has existed since the discovery of the first Be star $\gamma$ Cas by
Father Secchi in 1866.  The disk around $\alpha$ Arae is compatible
with a dense equatorial matter confined in the central region whereas
a polar wind is contributing along the rotational axis of the central
star. Between these two regions the density must be low enough to
reproduce the large visibility amplitudes obtained for two of the four
VLTI baselines.
 

Thanks to these first spectrally resolved interferometric measurements
of a Be star at 2 $\mu$m we were able to propose a possible scenario
for the Be star $\alpha$ Arae circumstellar environment which
consists of a thin disk and polar enhanced winds that is successfully
modeled with the SIMECA code (see Fig.~\ref{fig:map_continu}).

Using differential visibility amplitudes and phases across the
Br$\gamma$ line, Meilland et al.\cite{2006meilland} detected an asymmetry
in the circumstellar structure around $\kappa$ Canis Majoris. However,
$\kappa$ CMa is difficult to fit within the classical scenario for Be
stars, i.e. a fast rotating B star close to its breakup velocity
surrounded by a Keplerian circumstellar disk with an enhanced polar
wind. We found that $\kappa$ CMa does not seem  to be a critical rotator,
the rotation law within the disk is not Keplerian and the detected
asymmetry seems to be hardly explained within the one-armed viscous
disk framework.

\begin{figure}[t]
  \centering
  \includegraphics[width=\hsize]{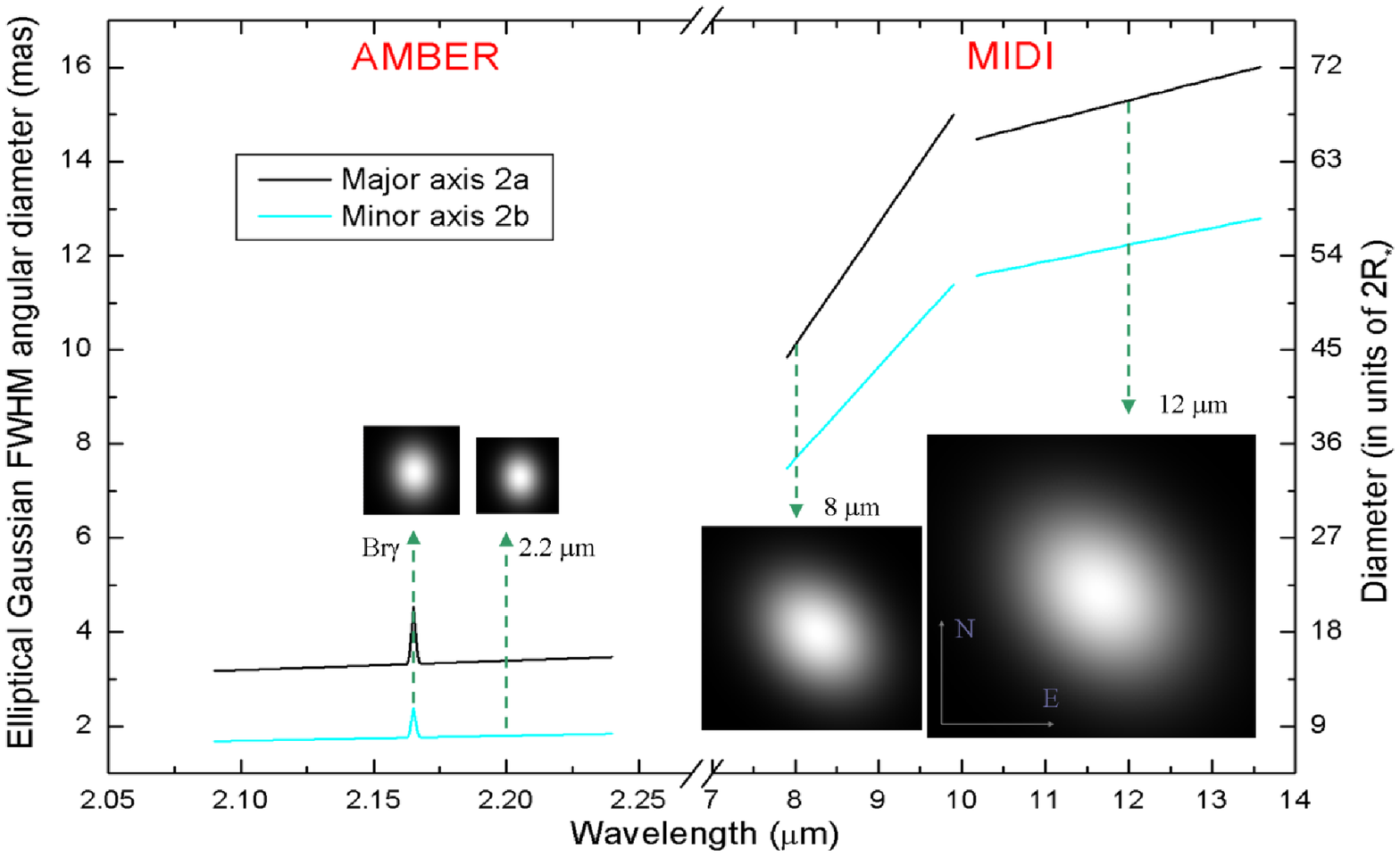}
  \caption{Wavelength-dependent sizes of the B[e] supergiant CPD-57
2874 derived from the fit of a chromatic elliptical Gaussian model
to the VLTI/AMBER and VLTI/MIDI visibilities. The scale in the
right is given in stellar diameters ($2R_*$), where the radius
$R_*$ is estimated to be $60\pm15\,\mathrm{R}_{\odot}$.}
\label{fig:cpd572874}
\end{figure}

The first high spatial and spectral observations of the circumstellar
envelope of CPD$-57\degr2874$, a B[e] supergiant, have been performed
with the VLTI.  Spectra, visibilities, and closure phase were obtained
using the beam-combiner instruments AMBER and MIDI
\cite{2005astro.ph.10735D}. In a first order analysis of the
wavelength-dependent size and geometry of the circumstellar envelope
of CPD-57{\degr}2874 we fitted the visibilities by an elliptical
Gaussian model with diameters varying linearly with wavelength.
Typical angular sizes of the major-axes are $3.4$~mas
($2.2\,\mu\mathrm{m}$ continuum from AMBER), $5.2$~mas (Br$\gamma$
emission line from AMBER), and $15$~mas ($12\,\mu\mathrm{m}$ continuum
from MIDI). The spectro-interferometric VLTI observations and data
analysis support the non-spherical, gaseous and dusty circumstellar
envelope paradigm for B[e] supergiants.

\section{Mass loss from massive stars}

$\eta$ Car is one of the most luminous ($L\sim4\times10^6\,L_\odot$)
and most massive ($M\sim100\,M_\odot$) unstable Luminous Blue
Variables suffering from an extremly high mass loss rate.
Spectroscopic studies of the Homunculus nebula showed that the wind of
$\eta$ Car is latitude-dependent.  Van Boeckel et
al.\cite{2003A&A...410L..37V} resolved the optically thick, aspheric
wind region with NIR interferometry using the VLTI/VINCI instrument.
This aspheric wind can be explained by models for line-driven winds
from luminous hot stars rotating near their critical speed.  The
models predict a higher wind speed and density along the polar axis
than in the equatorial plane.  A variety of observations suggest that
the central source of $\eta$ Car is a binary even if the
binary nature of the central object in $\eta$ Car is still a matter of
debate.


\begin{figure}[p]
\centering
  \includegraphics[height=0.8\textheight]{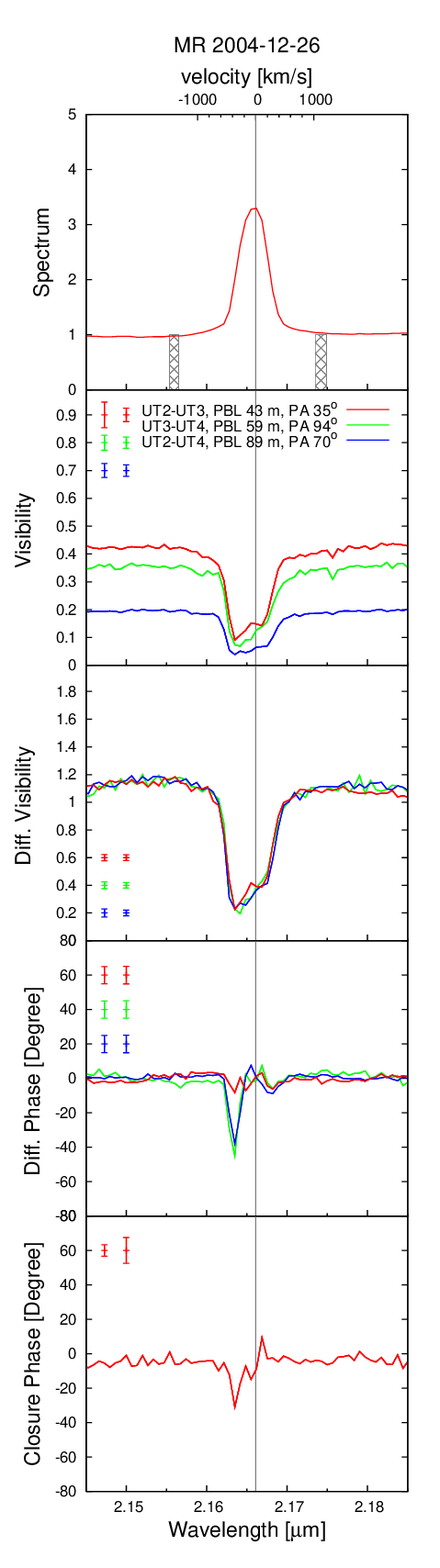}
  \caption{ AMBER observables derived from our $\eta$~Car data around
    the Br$\gamma$ line (MR mode, 2004~December~26; the HR mode and He
    I observations are described in Weigelt et al\cite{2006weigelt}).
    The top panel shows the spectrum as extracted from the
    interferometric channels, followed by the derived calibrated
    visibilities and the differential visibilities.  In the two panels
    at the bottom, the differential phase and the closure phase are
    presented.  The vertical grey line marks the rest-wavelength of
    Br$\gamma$ ($\lambda_{\rm vac}=2.1661\,\mu$m).  The left error
    bars correspond to the total error estimated for the continuum
    wavelength range, and the error bars towards the right visualizes
    the total error for the wavelength range within the line.  }
\label{fig:rawdata}
\end{figure}

Our AMBER observations of $\eta$ Car\cite{2006weigelt} were performed
with three Unit Telescopes and with spectral resolutions of $R=1,500$
(MR mode) and $R=12,000$ (HR mode) in Dec.\ 2004 ($\phi=0.268$) and
Feb.\ 2005 ($\phi=0.299$) in spectral windows around the HeI and
Br$\gamma$ emission lines at $\lambda=2.059$ and $2.166\,\mu$m,
respectively. From the measurements, we obtained spectra,
visibilities, differential visibilities, differential phases, and
closure phases.  With projected baseline lengths up to 89~m, an
angular resolution of $\sim$5 mas was achieved in the $K$ band.

In the $K$-band continuum, we resolved $\eta$ Car's optically thick
wind.  From a Gaussian fit of the $K$-band continuum visibilities in
the projected baseline range from 28--89~m, we obtained a FWHM
diameter of $4.0\pm0.2$ mas. When comparing the AMBER continuum
visibilities with the NLTE radiative transfer model from Hillier et
al.\cite{2001ApJ...553..837H}, we find very good agreement between the
model and observations. The best fit was obtained with a slightly
rescaled version of the original Hillier et
al.\cite{2001ApJ...553..837H} model, corresponding to an observed FWHM
diameter of 2.4~mas and a 50\% encircled-energy diameter of 4.3 mas at
$\lambda=2.17\,\mu$m.  Taking the different FOVs into account, we
found good agreement between the AMBER measurements and previous
VLTI/VINCI observations of $\eta$ Car presented by van Boeckel et
al.\cite{2003A&A...410L..37V}. If we fit Hillier et
al.\cite{2001ApJ...553..837H} model visibilities to the observed AMBER
emission line visibilities, we obtain 50\% encircled-energy diameters
of 6.5 and 9.6~mas in HeI and the Br$\gamma$ emission lines,
respectively.


For both the Br$\gamma$ and the HeI emission lines, we measured
non-zero differential phases and non-zero closure phases within the
emission lines, indicating a complex, asymmetric object structure.  We
developed a physically motivated model\cite{2006weigelt}, which shows
that the asymmetries measured within the wings of the Br$\gamma$ line
with differential and closure phases are consistent with the geometry
expected for an aspherical, latitude-dependent stellar wind.

\section{Interacting binary in late stellar evolution}

The Wolf-Rayet and O star binary system $\gamma^2$ Velorum has been
investigated by AMBER. The goal was to understand better the wind of a
WR star in an interacting binary on this double line spectroscopic
system already observed with intensity interferometry
\cite{1970MNRAS.148..103H}.  Strong signals were obtained in all
interferometric measures, showing that the image of the system is
strongly dominated by two compact, unresolved sources around the two
stars location. The parameters of the binary have been accurately
derived and have been shown to be stable with regard to the different
approaches used to model the spectrum of the two components and to
evaluate the contribution of different kinds of circumstellar
material. In particular, Petrov et al.\cite{2006petrov} demonstrated that
the combination of differential visibility, differential phase and
closure phase as a function of $\lambda$ allows to obtain
independantly the angular separation vector and the spectra of the
components.  Therefore, they have a new measure of the distance at
$368_{-13}{^+38}$\,pc. This lies between the pre Hipparcos
spectro-photometric estimations (450\,pc) and the Hipparcos measure of
258\,pc, yielding one more reason to consider revision of Hipparcos
distances of complex sources close to the performance limit. A second
result of the work is a direct, model independent, measurement of the
spectrum of the WR component which opens the possibility of improved
modelling of this star. Whatever the fit
of the data with a binary system, the measures display, in tens of
spectral channels, a $5\sigma$ to $10\sigma$ residual which is the
signature of circumstellar material.  The dust contribution to the
continuum is limited to less than 5\% of the total flux. The residuals
exhibit spectral features which allow to speculate that they result
from interstellar gas, probably near the wind-wind collision zone,
contributing to the emission lines and also to the continuum with
free-free emission\cite{2006petrov}.

\section{Conclusion}

The first astrophysical results from AMBER have focused on
stellar wind structure, kinematics, and its interaction with dust
usually concentrated in a disk. Because AMBER has dramatically
increased the number of measures per baseline, this instrument brings
strong constraints on morphology and models despite a relatively poor
$(u, v)$ coverage for each object. 

From these initial measurements, we can deduce that there might be an
ubiquity of an equatorial disk of dust and/or gas with a
latitude dependant wind. We can wonder if this results from a
selection effect, a term bias or an actual breakthrough. For other
observing programs the reader can read the AMBER science
program\cite{2004SPIE.5491.1722M} described in last SPIE conference.

The VLTI is currently limited by vibrations and is under intensive
tests. Hopefully, the consolidation plan undertaken by ESO will soon
allow us to achieve the ultimate performance of AMBER.

\acknowledgments     
 
We would like to thanks all persons who made it possible to conceive,
build and install the AMBER instrument.


\bibliography{amber-spie}   
\bibliographystyle{spiebib}   

\end{document}